\documentclass{moriond}
\usepackage[utf8]{inputenc}

\def\be{\begin{equation}}
\def\ee{\end{equation}}
\def\bea{\begin{eqnarray}}
\def\eea{\end{eqnarray}}

\def\Nf{N_{\rm eff}}

\newcommand{\Photo}{}

\begin{document}
\vspace*{4cm}
\title{Cosmic Neutrinos and Other Light Relics}

\author{ Joel Meyers }

\address{Canadian Institute for Theoretical Astrophysics, Toronto, ON M5S 3H8, Canada}

\maketitle

\abstracts{
Cosmological measurements of the radiation density in the early universe can be used as a sensitive probe of physics beyond the standard model.  Observations of primordial light element abundances have long been used to place non-trivial constraints on models of new physics and to inform our understanding of the thermal history to the first few minutes of our present phase of expansion.  Precision measurements of the angular power spectrum of the cosmic microwave background temperature and polarization will drastically improve our measurement of the cosmic radiation density over the next decade.  These improved measurements will either uncover new physics or place much more stringent constraints on physics beyond the standard model, while pushing our understanding of the early universe to much earlier times.  
}

\section{Introduction}

The primordial light element abundances and the angular power spectrum of the cosmic microwave background (CMB) are sensitive to the total radiation density that was present in the early universe, usually parametrized through a quantity $\Nf$.  In the standard models of particle physics and cosmology, the value of $\Nf$ measures the total energy density of cosmic neutrinos.  In more general models, however, $\Nf$ receives contributions from all forms of relativistic species apart from photons present in the early universe.  This means that $\Nf$ can be used as a sensitive probe of physics beyond the standard model.  The next generation of CMB experiments will drastically improve our constraints on $\Nf$, opening a characteristically new regime in the cosmological search for new physics. 

The quantity $\Nf$ is typically defined as the ratio of the energy density of all dark radiation (that is all radiation except photons) to that of photons ~\cite{Weinberg:2008zzc}
\begin{equation}
	\Nf = \frac{8}{7}\left(\frac{11}{4}\right)^{4/3}\frac{\rho_X}{\rho_\gamma} \, .
\end{equation}
The standard model predicts $\Nf = 3.046$ ~\cite{Mangano:2005cc}. %, though differing numerical approximations lead to predictions which vary at the level of about $10^{-3}$ ~\cite{Grohs:2015tfy}.

\section{Observational Signatures}

\subsection{Primordial Light Element Abundances}

The process by which light elements form in the early universe, known as big bang nucleosynthesis (BBN), is affected by the presence of dark radiation due to its impact on the expansion rate.  For example, an increase in $\Nf$ tends to produce a higher helium-4 abundance.  For higher radiation density, the expansion rate is higher, which implies that the temperature drops more quickly with time.  This results in a higher neutron-to-proton freeze-out ratio and less time for free neutron decay. Thus, there remains a higher number density of neutrons surviving when the universe cools sufficiently for the formation of helium nuclei, thereby producing more helium-4 than in the standard scenario~\cite{Weinberg:2008zzc,Agashe:2014kda,Cyburt:2015mya}.

Astrophysical measurements of the primordial helium-4 abundance have for a long time provided the best constraint on $\Nf$, though recently improved measurements of the primordial deuterium abundance have become competitive with those of helium-4~\cite{Cooke:2013cba,Cyburt:2015mya}.  The current best measurement of $\Nf$ from primordial abundances is~\cite{Cyburt:2015mya}
\begin{equation}
	\Nf^\mathrm{BBN} = 3.28 \pm 0.28 \, ,
\end{equation}
which is fully consistent with the predictions of the standard model.

%In addition to being sensitive to the total radiation density, the formation of primordial light elements is also weakly dependent on the energy density and distribution function of active neutrinos, though this is subdominant to the dependence $\Nf$ for small non-thermal distortions of the neutrino spectrum~\cite{Serpico:2004gx}.

\subsection{Cosmic Microwave Background}

Dark radiation impacts the angular power spectrum of the CMB in a number of ways.  The most prominent observational signature is the effect of increased damping in the presence of higher radiation density~\cite{Hou:2011ec}.  Perturbations in free-streaming radiation (such as cosmic neutrinos) also lead to shifts in the amplitude and phase of the acoustic peaks which approach constants at small angular scales~\cite{Bashinsky:2003tk,Baumann:2015rya}.  The phase shift is of particular interest, since it is not degenerate with other cosmological parameters and  will become increasingly important in driving future constraints on $\Nf$~\cite{Baumann:2015rya}.  Furthermore, since the phase shift is only sourced by free-streaming radiation, measurements of the CMB can also be used to distinguish the nature of dark radiation~\cite{Bell:2005dr,Friedland:2007vv,Chacko:2015noa,Baumann:2015rya}.  The phase shift due to cosmic neutrinos has recently been isolated and detected in the Planck temperature data~\cite{Follin:2015hya}.

Measurements from the Planck satellite using both temperature and polarization currently provide the best constraint on $\Nf$, giving~\cite{Ade:2015xua}
\begin{equation}
	\Nf^\mathrm{CMB} = 3.04 \pm 0.18 \, .
\end{equation}
We see that this measurement is consistent with the prediction of the standard model and also with the measurements of primordial abundances.

Future measurements of the CMB, in particular from a ground-based CMB Stage-IV experiment, are expected to improve on the current constraints on $\Nf$ by about an order of magnitude~\cite{Abazajian:2013oma,Wu:2014hta}.  The phase shift of the acoustic peaks will play an increasingly important role in these constraints~\cite{Baumann:2015rya}.  The $E$-mode polarization power spectrum has sharper acoustic peaks than that of the temperature power spectrum, thus yielding a more accurate measurement of the phase shift and $\Nf$.  Since gravitational lensing of the CMB tends to smooth out acoustic peaks, delensing which sharpens peaks can improve the constraints on $\Nf$~\cite{Baumann:2015rya,delensing}.

\subsection{Complementarity}

Measurements of $\Nf$ from the CMB are now more precise than those from primordial abundance observations, though these should be viewed as complementary rather than competing probes.  The two measurements are sensitive to different aspects of the radiation content of the universe (since for example the primordial abundances depend weakly on the distribution function of active neutrinos~\cite{Serpico:2004gx}), and so combining these measurements provides a more comprehensive picture than either alone.  Furthermore, these two measurements are sensitive to the radiation energy density at different times in the cosmic history, thereby allowing constraints on the evolution of the radiation density and the thermal history of the universe~\cite{Fischler:2010xz,Cadamuro:2010cz,Menestrina:2011mz,Hooper:2011aj,Millea:2015qra}.

The CMB is also directly sensitive to the primordial helium abundance through its effect on the damping tail.  Such constraints from CMB Stage-IV will be more precise and much less sensitive to astrophysical systematics than the current best observations of primordial helium~\cite{Wu:2014hta,Baumann:2015rya}.

\section{Theoretical Targets}

Precise measurements of $\Nf$ naturally place constraints on many extensions of standard model physics.  This includes for example gravitational waves~\cite{Boyle:2007zx,Stewart:2007fu,Meerburg:2015zua}, dark photons~\cite{Ackerman:mha,Kaplan:2011yj,CyrRacine:2012fz}, sterile neutrinos~\cite{Abazajian:2001nj,Strumia:2006db,Boyarsky:2009ix}, and many more~\cite{Cadamuro:2010cz,Weinberg:2013kea}.

Perhaps one of the most compelling motivations for studying $\Nf$ is that all light thermal relics contribute to $\Nf$ at a level which is determined by the spin and decoupling temperature alone~\cite{Brust:2013ova,Chacko:2015noa}.  The minimum contribution from a thermal relic which decoupled after the QCD phase transition is roughly $\Delta\Nf\sim 0.3$ which is the regime currently probed by the Planck observations.  The measurements of CMB Stage-IV will be sensitive to light relics which decoupled at much earlier times and higher temperatures.  Light thermal relics with essentially arbitrarily high decoupling temperature always~\footnote{This conclusion can be avoided if a large number of dark sector species decay into standard model particles after the decoupling of the dark radiation, thereby diluting its relic abundance.} produce $\Delta \Nf \geq 0.027$~\cite{Brust:2013ova,Chacko:2015noa}.  A CMB experiment which reaches this level of sensitivity will therefore either detect new physics, or rule out the existence of new light thermal relics, a conclusion which would have extremely far reaching consequences for particle physics~\cite{Baumann:2016wac}.

Measurements of $\Nf$ therefore allow the CMB to be used as a window onto particle physics at energies much higher than are accessible by other means.  The next decade of CMB observations will thus be an extremely exciting period for cosmologists and particle physicists alike.

\section*{Acknowledgments}
Thanks to Daniel Baumann, Dan Green, Alex van Engelen, and Benjamin Wallisch for their contributions to the research which forms the basis of this article.

\section*{References}
\bibliography{neff} 

\begin{thebibliography}{10}

\bibitem{Weinberg:2008zzc}
Steven Weinberg.
\newblock {\em {Cosmology}}.
\newblock 2008.

\bibitem{Mangano:2005cc}
Gianpiero Mangano, Gennaro Miele, Sergio Pastor, Teguayco Pinto, Ofelia
  Pisanti, and Pasquale~D. Serpico.
\newblock {Relic neutrino decoupling including flavor oscillations}.
\newblock {\em Nucl. Phys.}, B729:221--234, 2005.

\bibitem{Agashe:2014kda}
K.~A. Olive et~al.
\newblock {Review of Particle Physics}.
\newblock {\em Chin. Phys.}, C38:090001, 2014.

\bibitem{Cyburt:2015mya}
Richard~H. Cyburt, Brian~D. Fields, Keith~A. Olive, and Tsung-Han Yeh.
\newblock {Big Bang Nucleosynthesis: 2015}.
\newblock {\em Rev. Mod. Phys.}, 88:015004, 2016.

\bibitem{Cooke:2013cba}
Ryan Cooke, Max Pettini, Regina~A. Jorgenson, Michael~T. Murphy, and Charles~C.
  Steidel.
\newblock {Precision measures of the primordial abundance of deuterium}.
\newblock {\em Astrophys. J.}, 781(1):31, 2014.

\bibitem{Hou:2011ec}
Zhen Hou, Ryan Keisler, Lloyd Knox, Marius Millea, and Christian Reichardt.
\newblock {How Massless Neutrinos Affect the Cosmic Microwave Background
  Damping Tail}.
\newblock {\em Phys. Rev.}, D87:083008, 2013.

\bibitem{Bashinsky:2003tk}
Sergei Bashinsky and Uros Seljak.
\newblock {Neutrino perturbations in CMB anisotropy and matter clustering}.
\newblock {\em Phys. Rev.}, D69:083002, 2004.

\bibitem{Baumann:2015rya}
Daniel Baumann, Daniel Green, Joel Meyers, and Benjamin Wallisch.
\newblock {Phases of New Physics in the CMB}.
\newblock {\em JCAP}, 1601:007, 2016.

\bibitem{Bell:2005dr}
Nicole~F. Bell, Elena Pierpaoli, and Kris Sigurdson.
\newblock {Cosmological signatures of interacting neutrinos}.
\newblock {\em Phys. Rev.}, D73:063523, 2006.

\bibitem{Friedland:2007vv}
Alexander Friedland, Kathryn~M. Zurek, and Sergei Bashinsky.
\newblock {Constraining Models of Neutrino Mass and Neutrino Interactions with
  the Planck Satellite}.
\newblock 2007.

\bibitem{Chacko:2015noa}
Zackaria Chacko, Yanou Cui, Sungwoo Hong, and Takemichi Okui.
\newblock {Hidden dark matter sector, dark radiation, and the CMB}.
\newblock {\em Phys. Rev.}, D92:055033, 2015.

\bibitem{Follin:2015hya}
Brent Follin, Lloyd Knox, Marius Millea, and Zhen Pan.
\newblock {First Detection of the Acoustic Oscillation Phase Shift Expected
  from the Cosmic Neutrino Background}.
\newblock {\em Phys. Rev. Lett.}, 115(9):091301, 2015.

\bibitem{Ade:2015xua}
P.~A.~R. Ade et~al.
\newblock {Planck 2015 results. XIII. Cosmological parameters}.
\newblock 2015.

\bibitem{Abazajian:2013oma}
K.~N. Abazajian et~al.
\newblock {Neutrino Physics from the Cosmic Microwave Background and Large
  Scale Structure}.
\newblock {\em Astropart. Phys.}, 63:66--80, 2015.

\bibitem{Wu:2014hta}
W.~L.~K. Wu, J.~Errard, C.~Dvorkin, C.~L. Kuo, A.~T. Lee, P.~McDonald,
  A.~Slosar, and O.~Zahn.
\newblock {A Guide to Designing Future Ground-based Cosmic Microwave Background
  Experiments}.
\newblock {\em Astrophys. J.}, 788:138, 2014.

\bibitem{delensing}
Daniel Green, Joel Meyers, and Alexander van Engelen.
\newblock {An All-Orders Approach to Delensing}, Forthcoming, 2016.

\bibitem{Serpico:2004gx}
Pasquale~Dario Serpico, S.~Esposito, F.~Iocco, G.~Mangano, G.~Miele, and
  O.~Pisanti.
\newblock {Nuclear reaction network for primordial nucleosynthesis: A Detailed
  analysis of rates, uncertainties and light nuclei yields}.
\newblock {\em JCAP}, 0412:010, 2004.

\bibitem{Fischler:2010xz}
Willy Fischler and Joel Meyers.
\newblock {Dark Radiation Emerging After Big Bang Nucleosynthesis?}
\newblock {\em Phys. Rev.}, D83:063520, 2011.

\bibitem{Cadamuro:2010cz}
Davide Cadamuro, Steen Hannestad, Georg Raffelt, and Javier Redondo.
\newblock {Cosmological bounds on sub-MeV mass axions}.
\newblock {\em JCAP}, 1102:003, 2011.

\bibitem{Menestrina:2011mz}
Justin~L. Menestrina and Robert~J. Scherrer.
\newblock {Dark Radiation from Particle Decays during Big Bang
  Nucleosynthesis}.
\newblock {\em Phys. Rev.}, D85:047301, 2012.

\bibitem{Hooper:2011aj}
Dan Hooper, Farinaldo~S. Queiroz, and Nickolay~Y. Gnedin.
\newblock {Non-Thermal Dark Matter Mimicking An Additional Neutrino Species In
  The Early Universe}.
\newblock {\em Phys. Rev.}, D85:063513, 2012.

\bibitem{Millea:2015qra}
Marius Millea, Lloyd Knox, and Brian Fields.
\newblock {New Bounds for Axions and Axion-Like Particles with keV-GeV Masses}.
\newblock {\em Phys. Rev.}, D92(2):023010, 2015.

\bibitem{Boyle:2007zx}
Latham~A. Boyle and Alessandra Buonanno.
\newblock {Relating gravitational wave constraints from primordial
  nucleosynthesis, pulsar timing, laser interferometers, and the CMB:
  Implications for the early Universe}.
\newblock {\em Phys. Rev.}, D78:043531, 2008.

\bibitem{Stewart:2007fu}
Andrew Stewart and Robert Brandenberger.
\newblock {Observational Constraints on Theories with a Blue Spectrum of Tensor
  Modes}.
\newblock {\em JCAP}, 0808:012, 2008.

\bibitem{Meerburg:2015zua}
P.~Daniel Meerburg, Renee Hlozek, Boryana Hadzhiyska, and Joel Meyers.
\newblock {Multiwavelength constraints on the inflationary consistency
  relation}.
\newblock {\em Phys. Rev.}, D91(10):103505, 2015.

\bibitem{Ackerman:mha}
Lotty Ackerman, Matthew~R. Buckley, Sean~M. Carroll, and Marc Kamionkowski.
\newblock {Dark Matter and Dark Radiation}.
\newblock {\em Phys. Rev.}, D79:023519, 2009.
\newblock [,277(2008)].

\bibitem{Kaplan:2011yj}
David~E. Kaplan, Gordan~Z. Krnjaic, Keith~R. Rehermann, and Christopher~M.
  Wells.
\newblock {Dark Atoms: Asymmetry and Direct Detection}.
\newblock {\em JCAP}, 1110:011, 2011.

\bibitem{CyrRacine:2012fz}
Francis-Yan Cyr-Racine and Kris Sigurdson.
\newblock {Cosmology of atomic dark matter}.
\newblock {\em Phys. Rev.}, D87(10):103515, 2013.

\bibitem{Abazajian:2001nj}
Kevork Abazajian, George~M. Fuller, and Mitesh Patel.
\newblock {Sterile neutrino hot, warm, and cold dark matter}.
\newblock {\em Phys. Rev.}, D64:023501, 2001.

\bibitem{Strumia:2006db}
Alessandro Strumia and Francesco Vissani.
\newblock {Neutrino masses and mixings and...}
\newblock 2006.

\bibitem{Boyarsky:2009ix}
Alexey Boyarsky, Oleg Ruchayskiy, and Mikhail Shaposhnikov.
\newblock {The Role of sterile neutrinos in cosmology and astrophysics}.
\newblock {\em Ann. Rev. Nucl. Part. Sci.}, 59:191--214, 2009.

\bibitem{Weinberg:2013kea}
Steven Weinberg.
\newblock {Goldstone Bosons as Fractional Cosmic Neutrinos}.
\newblock {\em Phys. Rev. Lett.}, 110(24):241301, 2013.

\bibitem{Brust:2013ova}
Christopher Brust, David~E. Kaplan, and Matthew~T. Walters.
\newblock {New Light Species and the CMB}.
\newblock {\em JHEP}, 12:058, 2013.

\bibitem{Baumann:2016wac}
Daniel Baumann, Daniel Green, and Benjamin Wallisch.
\newblock {A New Target for Cosmic Axion Searches}.
\newblock 2016.

\end{thebibliography}
\end{document}